\documentclass[10pt,a4paper]{article}
\usepackage{graphics}

\def\m#1{$#1$}
\def\tr{\;{\rm tr}\;}

\newcommand{\DOE}{This work was supported in part by U.S.Department
    of Energy grant No. DE--FG02-91ER40685}

\newcommand{\beq}{\begin{equation}}
\newcommand{\eeq}{\end{equation}}
\newcommand{\dd}[2]{\frac {\partial #1}{\partial #2}}
\newcommand{\pdr}{\partial}
\newcommand{\beqs}{\begin{eqnarray}}
\newcommand{\eeqs}{\end{eqnarray}}

\newcommand{\half}{\frac{1}{2}}

\newcommand{\ov}[1]{\frac{1}{#1}}

\newcommand{\Nsq}{{1 \over N^2}}
\newcommand{\Ntr}{{{\rm tr} \over N}}
\newcommand{\La}{\Lambda}
\newcommand{\la}{\lambda}

\newcommand{\calS}{{\cal S}}

\newcommand{\grad}{{\bf \nabla}}

\newcommand{\p}{{\bf p}}

\newtheorem{sutra}{}

\newtheorem{bhashya}{}[sutra]

\begin{document}
\bibliographystyle{h-physrev}
\input{epsf}

\title{Collective potential for large-$N$ hamiltonian matrix
models and free Fisher information}
\author{A.Agarwal\thanks{abhishek@pas.rochester.edu} \\
\and
 L.Akant\thanks{akant@pas.rochester.edu}\\
\and
 G.S.Krishnaswami\thanks{govind@pas.rochester.edu}\\
\and
 S.G.Rajeev\thanks{rajeev@pas.rochester.edu} \\
University of Rochester. Dept of Physics and Astronomy. \\
Rochester. NY - 14627}
\maketitle

\begin{abstract}
We formulate the planar `large N limit' of matrix models with a
continuously infinite number of matrices directly in terms of \m{U(N)} invariant
variables. Non-commutative probability theory, is found to be a good
 language to describe this formulation. The change of variables from
 matrix elements to invariants  induces an extra term in the
 hamiltonian,which is crucual in determining the ground state. We find that
 this collective potential  has a natural meaning in terms of non-commutative probability
 theory:it is the `free Fisher information' discovered by Voiculescu.
 This  formulation allows us
to find a variational principle for the classical theory described
by such large N limits.   We then use the variational principle to
study models more complex than the one describing the quantum
mechanics of a single hermitian matrix (i.e., go beyond the so
called $D=1$ barrier). We carry out approximate variational
calculations for a few models and find excellent agreement with
known results where such comparisons are possible. We also
discover a lower bound for the ground state by using the
non-commutative analogue of the Cramer-Rao inequality.
\end{abstract}

\pagebreak
\section{Introduction}
The purpose of the present paper is to find the natural
mathematical language for describing the `large N limit' of matrix
models involving a continuously infinite number of matrices (in
the Hamiltonian picture) e.g matrix quantum mechanics and matrix
field theories. We find that non-commutative probability theory
offers such a language, which after suitable generalizations can
be adopted for the purposes of formulating the large N limit.
Recasting continuum matrix models in this language allows us to
find a variational principle for calculating the gauge invariant
observables of such models. The advantage of the variational
method being that we can use it to make well controlled
approximations when exact solutions are no longer possible.

No apologies are really necessary for the study of continuum matrix models
as they are ubiquitous in the world of both gauge theories and strings.
QCD,  where the gluonic field $(A_\mu )$ belongs to the
adjoint representation of the structure group, is perhaps the most natural
example of a field theory where the dynamical degrees of freedom are
matrix valued. It is believed that the planar large N limit is the correct
simplification to consider while trying to understand non-perturbative
phenomena such as confinement, understanding which in the physically
interesting case of four spacetime dimensions remains one of the major
unsolved problems of theoretical physics.  QCD when dimensionally reduced
to one dimension is known to provide a regularization of the theory of the
classical membrane, and the generalization of this method to the case of
the supermembrane provides one with the  Hamiltonian which is conjectured
to be a description of M-theory in the infinite momentum frame \cite{
BFSS}. Other
examples of one dimensional matrix models include those with  a
single matrix (with polynomial self interactions) depending continuously on
a parameter (which is usually taken to be physical or stochastic time).
Such models are known to provide explicit realizations
of string field theories in one dimensional target spaces \cite{Stringf1,
stringf2}.

In a previous paper\cite{entropy}  we studied `static' matrix models, involving the integral
over a finite number of hermitean matrices:
\beq
    \int e^{N\tr S(A)} {1\over N}\tr A_{i_1}\cdots A_{i_n} dA.
\eeq It is known that the theory of `free random variables'
\cite{Voicbook}has a lot to do with the large N limit of these
models \cite{Gross-Gopakumar, Doug1, Doug2, entropy}. The
mathematician's notion of `freeness' \cite{Voic1, Voic2, Voic3,
Voic4} and the physicist's picture of the planar large N limit are
intimately related. The large N limit of matrix models was
formulated in terms of a variational principle. There is an
anamolous term in this variational principle, which we showed, has
a simple meaning in terms of non-commutative probability theory:
it is the free entropy of Voiculescu.

In the present paper we will extend our study to `dynamical'
matrix models, determined by  hamiltonians such as \beq
-K_{ij}{\pdr^2\over \pdr A_{ia}^b\pdr A_{jb}^a}+\tr \tilde V(A).
\eeq In the  path integral formulation of quantum mechanics this
leads to integrals
 such as
\beq
\int e^{-\int\left[K_{ij}\tr\dot A_i\dot A_j+\tilde V(A) \right]dt}\tr A_{i_1}(t_1)\cdots A_{i_n}(t_n) {\cal D }A.
\eeq

Starting from the quantum theory at finite `N', there has been
considerable progress in extracting the large N limit, which is
described by a classical Hamiltonian. The `Collective Field'
formalism of Jevicki and Sakita \cite{Collective1, Collective2,
Collective3,Collective4, Collective5} and the `coherent state'
picture of Yaffe \cite{Yaffe1} stand out in this area.
The idea is to change variables from the matrix elements to the invariants of
the matrices, because quantum fluctuations in these invariants become small in
the large N limit. Thus the theory when expressed in terms of invariants should
be a classical theory.   One of the
key lessons of their work is that this large N classical limit  differs
in many deep ways from the more familiar classical limit as \m{\hbar\to 0}. Most
striking is the appearance of certain universal terms in the gauge
invariant formulation of the Large N theory (e.g. The $\phi ^3$
term in the collective potential of Jevicki and Sakita).

To understand the origin of these terms, let us consider a toy model from
elementary quantum mechanics:
a rotation invariant system whose configuration space is \m{R^N} and has
 hamiltonian operator
\beq
    \tilde H =-{\hbar^2}{\pdr^2 \over \pdr x^i\pdr x^i}+\tilde V(r), \quad r=\surd [x^ix^i].
\eeq If we consider only wavefunctions that are rotation invariant
( zero angular momentum), the angular variables can be eliminated
and the hamiltonian can be expressed as a differential operator in
the invariant variable \m{r}. The simplest way to perform this
change of variables is to note that \beq \int
\left[\hbar^2\pdr_i\psi^*\pdr_i\psi + \tilde
V(r)|\psi|^2\right]d^Nx=C_N\int\left[\hbar^2|\psi'|^2+\tilde
V(r)|\psi|^2\right]r^{N-1}dr \eeq where  \m{C_N r^{N-1}} is the
area of the sphere of radius \m{r}. This factor
 is the jacobian of the non-linear change of variables \m{x^i\to r}. Even
 if we make the
change of variables to the `radial wavefunction'
\m{\Psi(r)=r^{N-1\over 2}\psi(r)} to absorb this jacobian factor, there will be
some derivatives of the jacobian that survive. After an integration by parts,
we get
\beq
    C_N\int\left[\hbar^2|\Psi'|^2+\left\{\hbar^2{(N-1)(N-3)\over
    4r^2}+\tilde V(r)\right\}|\Psi|^2\right]dr
\eeq
Thus, the radial hamiltonian is not just \m{-\hbar^2{\pdr^2\over \pdr
r^2}+V(r)}, but is instead,
\beq
\tilde H=-\hbar^2{\pdr^2\over \pdr
r^2}+{\hbar^2}{(N-1)(N-3)\over 4r^2}+\tilde V(r).
\eeq
The extra term  is a sort of `centrifugal potential' arising from
differentiation  of the volume of the sphere of radius \m{r}. It is
at first surprising that such a centrifugal potential is present even when the
angular momentum is zero: a phenomenon with no analogue in the classical theory.
Indeed, this term vanishes as \m{\hbar\to 0}. This term is universal in the
sense that it is independent of the choice of potential: it arises from
rewriting the kinetic energy in radial variables.

To complete the story of this toy model, we see how to take the large N limit.
In this limit we should expect that
\beq
\rho^2={1\over N}x_ix_i
\eeq
should have small fluctuations. (This is a kind of central limit theorem.) Thus
expressed in these variables, the theory should approach a classical limit. This
classical limit will be very different from the more obvious limit \m{\hbar\to
0}. In particular, the ground state can be very different.
The hamiltonian is
\beq
\tilde H=-{\hbar^2\over N}{\pdr^2\over \pdr
\rho^2}+{\hbar^2}{(N-1)(N-3)\over 4N\rho^2}+\tilde V(N^\half \rho).
\eeq
If we define \m{H={\tilde H\over N}}, \m{V(r)={1\over N}\tilde V(N^\half \rho)}
we get
\beq
H=-{\hbar^2\over N^2}{\pdr^2\over \pdr
\rho^2}+{\hbar^2}{(N-1)(N-3)\over 4N^2\rho^2}+V(\rho).
\eeq
If we define
\beq
    \pi=-{i\over N}{\pdr \over \pdr \rho }
\eeq
we get the commutation relations
\beq
    [\pi,\rho]=-{i\over N}
\eeq and the hamiltonian \beq
H=-\hbar^2\pi^2+V_{coll}(\rho)+V(\rho), \eeq where
\m{V_{coll}={\hbar^2 (N-1)(N-3)\over N^2}{1\over 4 \rho^2}} Now we
can take the limit as \m{N\to \infty} keeping \m{V(\rho), \hbar}
fixed. The collective potential tends to \beq V_{coll}(\rho)\to
{\hbar^2\over 4}\Phi(\rho),\quad \Phi(\rho)={1\over \rho^2}. \eeq
The quantum fluctuations in \m{\rho} are indeed small because the
commutator \m{[\pi,\rho]\sim {1\over N}}. In the limit we get a
classical hamiltonian \beq
    H=\hbar^2p^2+{\hbar^2\over 4}\Phi(\rho)+V(\rho)
\eeq with the canonical Poisson bracket relations \beq
\{\rho,p\}=1. \eeq The ground state of this system is given by
\m{p=0, \rho=\rho_0} where \m{\rho_0} is the minimum of \beq
{\hbar^2\over 4}\Phi(\rho)+V(\rho). \eeq In the conventional
classical limit \m{\hbar\to 0} keeping \m{N} fixed, the ground
state would have been at the minimum of \m{V(\rho)}. For example,
if \m{V(\rho)=\half \omega^2 \rho^2}, it has  minimum at
\m{\rho=0} with vanishing ground state energy. On the other hand,
the ground state of the large $N$ limit is concentrated at \beq
\rho_0=\left[\hbar\over \sqrt{2} \omega\right]^\half \eeq with
$E_{gs} = {\hbar \omega \over \sqrt{2}}$. Thus the large $N$ limit
better captures the zero point fluctuations of this system.

The passage from matrix elements to invariants of a collection of matrices is
 analogous to the above change of variables from cartesian to radial co-ordinates. There is a
 jacobian associated with this nonlinear change. Transforming to `radial
  wavefunctions' to absorb this will again  induce an extra term in the
  potential, involving  the derivative of the jacobian.
These universal terms  enable the Large $N$ limit to capture
features of the theory that the usual perturbation expansion
misses. They are crucial in determining the ground state of the
matrix model.  Is it possible to understand them directly in the
large \m{N} limit?

In this paper we show that this universal term has a natural
meaning in the world of non-commutative probability theory: it is
the free analogue of the Fisher information obtained by Voiculescu
\cite{Voic1,Voic2,Voic3,Voic4}. This continues our earlier work
\cite{entropy}  where it is shown that the jacobian of the change
of variables to invariants  is just the exponential of free
entropy. Information is the square of the gradient of entropy,
which gives a rough explanation of our present result.

We first motivate the
connection between free probability theory and  large N  matrix
models in the context of static models, explaining some concepts from non-commutative probability
theory that appear not to be widely known  in the physics literature.
Then we  carry out this reformulation for
the
problem of matrix quantum mechanics (or field theory at D(dimension of
spacetime) = 1). This allows us to recover the results of Jevicki-Sakita
 \cite{Collective1, Collective2,
Collective3,Collective4, Collective5}.
Certain points of view on non-commutative probability theory generalize
beyond $D=1$, and we use them  to formulate a variational
principle for general matrix models.
 In the last part of the paper, we  apply this principle to some specific examples to
  get approximate answers for the ground state of  matrix field
theories with \m{D>1}, where explicit solutions are not possible.

\section{Non-Commutative Probability and Static Matrix
Models:}

Recall \cite{Voicbook, entropy} that in  non-commutative probability theory, the
analogue of a probability distribution is a rule that gives an expectation value
of a polynomial in the basic random variables. More precisely,
consider  the free associative algebra ${\cal T}_M= {\mathrm{C}}
\langle X_1,\cdots,X_M \rangle$ generated by the hermitian
operator valued random variables $X_i$ over the complex numbers.
We think of the elements of ${\cal T}_M$ as formal power series\footnote{
We continue the notation in \cite{entropy}. An upper case letter such as \m{I}
denotes a string of indices \m{i_1\cdots i_n}. Repeated multi-indices are
summed; \m{\phi} denotes the empty string.}
$f(X) = f^{i_1 \cdots i_n} X_{i_1\cdots i_n} = f^I X_I$. The joint
probability distribution of the $X_i$ in a tracial state is
specified by the expectation values $G_I = <X_I>$, which are
cyclically symmetric  tensors (`moments') satisfying
 \beq G_{\phi} =1,~~~ G_I^* = G_{\bar I}, ~~~ G_{IJ} {f^{\bar I}}^* f^J
 \geq 0 ~ {\rm for~any~polynomial~} f(X)
 \eeq
where $\bar I$ is the string with the opposite order. The last
condition is the positivity of the `Hankel matrix'  $G_{I;J} =
G_{IJ}$. It is used to construct a metric on the space of
probability distributions. $G^{I;J} G_{JK} = \delta^I_K$ denotes
the inverse.

If there were only one such random matrix \m{X} ( i.e., \m{M=1}), it would be
possible to introduce a probability density function on the real line
\m{\rho_G(x)} such that
\beq
    G_n=\int \rho_G(x)x^n dx.
\eeq
Many interesting quantities (such as entropy or information) have simple
formulas in terms of \m{\rho(x)}.

 The main complication in passing to
multimatrix models with \m{M>1} is that there is no longer any analogue of
\m{\rho(x)}. We have to think of non-commutative probability distributions
 indirectly in terms of the moments \m{G_I} or certain other cyclically
 symmetric tensors \m{S^I} which contain equivalent data.

 The two sets of
 tensors \m{S^I} and \m{G_I} are related by the `factorized Schwinger-Dyson
 equations':
 \beq
    S^{J_1 i J_2} G_{J_1 I J_2} + \eta^i_I =0,
 \eeq
 where $\eta^i_I = \delta_I^{I_1 i I_2} G_{I_1}G_{I_2}$.

 The motivation behind this definition is that the tensors \m{S^I} can be
 thought of as providing a `matrix model' for the non-commutative probability
 distribution:
 \beq
    G_I = \lim_{N \to \infty} {\int dA e^{N \tr S(A)}
    \Ntr A_I  \over \int dA e^{N \tr S(A)}}.
 \eeq
 Thus, we can think of \m{e^{N\tr S(A)}\over \int dA e^{N \tr S(A)} } as a
 (commutative) probability density function for the matrix elements of a set of
 \m{N\times N} matrices \m{A_i}. The quantity \m{\tr S(A)} is called the
 `action' of the matrix model in the physics literature. This extends to the
 realm of random variables the usual idea that non-commutative variables can be
 realized as matrices with complex numbers as entries. In a typical problem
 arising from quantum physics, the action \m{S(A)} is some simple polynomial and
we are required to calculate the moments arising from it: i.e., solve the
 Schwinger-Dyson equations.

  Infinitesimal variations of a non-commutative
  probability distribution  are vector
fields \m{L^i_I} on the space of probability distributions, with a natural action on
the moments given by,
\beq
    L^i_I G_J = \delta_J^{J_1 i J_2} G_{J_1 I J_2}.
 \eeq
Algebraically, $L^i_I$ are the derivations of the tensor algebra
${\cal T}_M$ corresponding to the
infinitesimal changes of variables $[\delta A]_j = \delta_j^i
A_I$.
These vector fields form a Lie algebra, with the following Lie bracket,
 \beq
   [L^i_I,L^j_J] = \delta ^{J_1iJ_2}_JL^j_{J_1IJ_2} -
\delta^{I_1jI_2}_{I}L^i_{I_1JI_2}.
 \eeq
For the case of the one matrix model, when the strings are labeled
by integers, this algebra reduces to the Virasoro algebra (without
the central term).

The quantity  $\eta^i_I$ appearing in the Schwinger-Dyson
equation generates  the first cohomology
of the Lie algebra\cite{entropy, abhi1}: it can be viewed as the variation
\m{\eta^i_I(G)=L^i_I\chi(G)} of some function \m{\chi(G)} which, however cannot be expressed as a formal power
series in the moments. Thus it is possible to regard the Schwinger-Dyson
equation as the condition for a function
\beq
\Omega(G)=S^IG_I+\chi(G)
\eeq
to be stationary: \m{L^i_I\Omega=0}. This variational principle is useful
because it allows for approximate solution of the SD equations: it is rare that
they can be solved exactly.

This quantity \m{\chi(G)} has a natural meaning in non-commutative probability
theory: up to an  additive constant,
$\chi(G)$ is the free entropy in the sense of Voiculescu
\cite{Voic1}. It is  the logarithm of volume of the set of $N\times N$
hermitian matrices whose moments are $G_I$ in the $N\to \infty$
limit\cite{Mehta2}.

To  obtain a more explicit formula for \m{\chi}, we have to introduce a third way
of describing a non-commutative probability distribution:
 a transformation $\phi:A_i \mapsto \phi(A)_i = \phi^I_i
A_I$ that maps given distribution to a reference distribution,
i.e. $\phi: G \rightarrow \Gamma $. Under such a change of
variable, $\chi$ transforms as a 1-cocycle, it is incremented by
the expectation value of $\Nsq \log \det$ of the Jacobian of
$\phi$. We do not follow this direction here, as it is explained
in an earlier paper \cite{entropy}.

For a single hermitian random variable, there are simple explicit formulas for
entropy
 \beqs
    \chi(G) &=& {\cal P} \int~ dx dy~ \rho_G(x)~ \rho_G(y) ~\log|x-y| \cr
            &=& \int dx dy ~ \rho_\Gamma(x)~ \rho_\Gamma(y)
            \log \bigg( {\phi(x)-\phi(y) \over x-y} \bigg)+\chi(\Gamma)
 \eeqs
where $\rho_\Gamma(x) = \rho_G(\phi(x)) \phi'(x) $.

It is also worth recalling, that for the case of a single random matrix,
the Schwinger-Dyson equation can be written in a simpler form using  $\rho $:
\beq
    (H\rho)(x) = \ov{2\pi} S'(x).
 \eeq
 (This is called the Mehta-Dyson integral equation.)
Here $(H\rho)(x) = \ov{\pi} {\cal P} \int dy {\rho(y) \over x-y}$
is the Hilbert transform.

The expression $\eta^i_I = \delta_I^{I_1 i I_2} G_{I_1} G_{I_2}$
motivates the introduction of the operator $\partial$
\cite{Voic2}, the difference quotient gradient , whose components
are
 \beq
    \partial^j: {\cal T}_M \to {\cal T}_M \otimes {\cal T}_M;
    ~~~ \partial^j X_I = \delta_I^{I_1 j I_2} X_{I_1} \otimes
    X_{I_2}.
\eeq The partial difference quotient can be thought of as acting
on a product of matrices. It breaks the matrix product at each
occurrence of $X_j$, exposing two more matrix indices, thus
producing an element in the tensor product. For $M=1$, $\partial
\phi(x) = {\phi(x) - \phi(y) \over x-y}$. $\partial$ relates the
non-commutative Hilbert transform to the variation of entropy.

The analogue of the difference quotient, when acting on a trace of
a product of matrices rather than on a product of matrices, leads
to the cyclic gradient $\delta $ \cite{Voic3}:
 \beq
    \delta^i : {\cal T}_M \to {\cal T}_M; ~~~
    \delta^i X_I = \delta_I^{I_1 i I_2} X_{I_2 I_1}.
 \eeq
The ordering of multi-indices is explained by the matrix analogy.
For a single generator, the cyclic gradient reduces to the partial
derivative. The cyclic gradient of a polynomial is unchanged by
cyclic permutations of any of its coefficients: Let $S(X) = S^I
X_I$. Then
 \beq
    \delta^k S(X) = S^{I_1 k I_2} X_{I_2 I_1} = S^{(kK)} X_K
 \eeq
where $S^{(K)} = S^{k_1 \cdots k_m} + S^{k_m k_1 \cdots k_{m-1}} +
\cdots + S^{k_2 \cdots k_m k_1}$.

The formalism developed above allows us to get a formula for the
Hilbert transform of a collection of non-commutative random variables. We first
note that for a single hermitian variable, the Mehta-Dyson
equation $H\rho(x) = \ov{2\pi} S'(x)$ tells us that the Hilbert
transform is the derivative of the 1-matrix model action. For
several hermitian random variables, the Hilbert transform is a
vector $H^i(X)$. The appropriate generalization of the
`derivative' of the action $S^I A_I$, must depend only on the
cyclic part of the tensors $S^I$. The natural candidate is the
cyclic gradient
 \beq
    H^i(X) = \ov{2\pi} \delta^i S(X).
 \eeq
Using the definition of $\delta ^i$,
 \beq
    H^i(X) = \ov{2\pi} S^{K_1 i K_2} X_{K_2 K_1} = \ov{2\pi}
    S^{Ii} X_I
 \eeq
The formula for the Hilbert transform given above involves the
action $S(X)$. One can look for a formula for $H^i(X)$ in terms of
the moments of the matrix model. Using the factorized SD equations
in the form $S^{Ii} = \eta^i_J G^{I;J}$ we get
 \beq
    H^i(X) = \ov{2\pi} \eta^i_I G^{I;J} X_J = \ov{2\pi}
        G_{I_1} G_{I_2} G^{I_1 i I_2;J} X_J.
 \eeq
Obtaining the inverse Hankel matrix from the moments is about as
hard as getting the moments of a static matrix model given its
action. This formula may be written
 \beq
    L^i_I \chi = 2\pi \langle H^i(X) X_I \rangle
 \eeq
thereby relating the Hilbert transform to the variation of
entropy. Voiculescu used this as the defining property of the
Hilbert transform \cite{Voic2}.

Another quantity from the world of non-commutative probability theory that
is going to be important for our understanding of matrix models is the
notion of free Fisher information.
 Following Voiculescu, let us first consider Fisher information
for a single real random variable $f$:
 \beq
        \Phi(f) = \lim_{t \to 0^+} {\calS(f+\sqrt{t} g) - \calS(f) \over
        t}.
 \eeq
Here $g$ is Gaussian with unit covariance, $f$ and $g$ are
statistically independent and $\calS(f)$ is Boltzmann's entropy of
$f$. If $f$ has pdf $\rho(x)$, then
 \beq
    \calS(f) = -\int \rho(x) \log{\rho(x)}dx
 \eeq
Regarding $\Phi$ as the zero time derivative of a Gaussian process
starting at $f$, the heat equation gives
 \beq
    \Phi(f) = \int dx ~\rho(x) ~\bigg(\partial_x log{\rho(x)} \bigg)^2
 \eeq

For several real random variables $x_i$ with joint distribution
$\rho({\bf x})$, the Fisher information matrix
 is defined as
 \beq
    \Phi^{ij} = \int d {\bf x} \rho({\bf x}) \partial^i \log{\rho({\bf
x})}
                \partial^j \log{\rho({\bf x})}
 \eeq

By analogy, the free Fisher information of a
single hermitian random variable may be defined as \cite{Voic2} :
 \beq
    \Phi(X) = \lim_{t \to 0^+} {\chi(X+\sqrt{t} S) - \chi(X) \over
        t}.
 \eeq
where $S$ is the Wigner semi-circular distribution. If $X +
\sqrt{t} S$ has pdf $\rho(x,t)$, then $\chi(X+\sqrt{t}S) = {\cal
P} \int \rho(x,t) \rho(y,t) \log|x-y| dx dy$ is the free entropy.
The complex Burger equation
 \beq
    \dd{G}{t} + G \dd{G}{z} = 0
 \eeq
for the Cauchy transform $G(z,t) = \int {\rho(x,t) dx\over x-z},~
z \notin {\rm supp}(\rho)$ plays the role of the heat equation for
a semi-circular process and leads to
\beq
    \Phi(X) = 2\pi^2 \int ~((H\rho)(x))^2 ~ \rho(x) ~ dx
                =~ {2\pi^2 \over 3} \int~ dx~ (\rho(x))^3.
 \eeq

By analogy with the case of several real random variables, we
define the free Fisher information matrix for several non-commuting
random variables as
 \beq
        \Phi^{ij}(G) = 2\pi^2 \langle~ H^i(X) H^j(X) ~\rangle
 \eeq
$\Phi^{ij}(G)$ cannot be expressed solely in terms of the moments.
As with the Hilbert transform, we can express it in terms of the
$G_K$ and either the inverse Hankel matrix $G^{I;J}$ or the action
$S^I X_I$ whose large $N$ moments are $G_K$:
 \beqs
    \Phi^{ij}(G) &=& \half G^{I;J} \eta^i_I \eta^j_J =
        \half G^{I_1 i I_2; J_1 j J_2} G_{I_1} G_{I_2} G_{J_1} G_{J_2}
        \cr
    \Phi^{ij}(G) &=& \half S^{iI} S^{jJ} G_{IJ}.
    \label{formula-for-information}
 \eeqs
The equivalence of these two follows from the factorized SD
equations relating $S^I$ to $G_K$:
 \beq
    S^{Ii} = \eta^i_J G^{I;J}.
 \eeq
For a single generator this formula reduces to the familiar one
when we use the Mehta Dyson equation.
\beqs
    \Phi(G) &=& \half \sum_{m,n} (m+1)(n+1)S_{m+1} S_{n+1} G_{m+n} \cr
            &=& \half \langle~ (S'(x))^2 ~ \rangle \cr
            &=&  2\pi^2 \int~ ((H\rho)(x))^2 ~\rho(x) ~dx
 \eeqs

\section{Dynamical Matrix Models and Collective Field Theory}

Now let us turn to the main object of our interest, i.e. matrix
models in the Hamiltonian approach. To have a definite example in
mind we consider in some detail, the quantum mechanics of a
discrete number of hermitian matrices. In other words, we consider
dynamical multi-matrix models whose Hamiltonians are of the form
 \beq
    \tilde H  = -K_{ij}\half \dd{^2}{(A_i)^a_b \pdr (A_j)^b_a} + \tr[\tilde V^I A_I]
 \eeq
Here $A_i(t)$ are $N \times N$ hermitian matrices, whose time
evolution is described by the Hamiltonian above. The positive tensor \m{K_{ij}}
determining the kinetic energy is usually just \m{\hbar^2\delta_{ij}}. The ground state
energy is
 \beq
    E_{gs} = \lim_{N \to \infty} \ov{N^2}
        \min_{||\psi(A)||=1} \langle \psi | \tilde H | \psi \rangle
 \eeq
where the wave function $\psi(A)$ is invariant under the adjoint
action of the unitary group; $A_i \mapsto U A_i U^\dag ~ \forall
i$ and the inner product is with respect to the Lebesgue measure
on matrix elements.

The planar large N limit of the above theory is the limit \m{N\to
\infty} holding \m{V^I = N^{|I|-1}\tilde V^I} and \m{K_{ij}}
fixed. Moreover, only wavefunctions that are invariant under
\m{U(N)} are allowed. It will be convenient to make the rescaling
\m{A_i\to N A_i} so that \beq
    {\tilde H\over N^2}=H  = -{1\over N^4}K_{ij}\half \dd{^2}{(A_i)^a_b \pdr (A_j)^b_a} +
    {1\over N}\tr[ V^I A_I].
 \eeq

  It is  known that
in this limit, quantum fluctuations in the moments \m{G_I={1\over N}\tr A_I}
become small. Hence there ought to be some classical mechanical system that
is equivalent to this limiting case of the matrix model. The configuration space
of this classical mechanical system would be the space of probability
distributions. It is our task now to determine the hamiltonian and Poisson
brackets of this theory and use them to gain insight into the properties ( such
as the ground state) of the matrix model.

\subsection{The case of One Degree of Freedom}
 Such a
reformulation is  most transparent in  the case where the time
evolution of a single hermitian matrix is concerned,
\beq
        \tilde H = -\half \hbar^2 \dd{^2}{A^a_b \partial A^b_a} + \sum_n\tr \tilde V_n A^n.
 \eeq
If we rescale \m{A\to N A} this becomes \beq {\tilde H\over
N^2}=-\half{\hbar^2\over N^4}\dd{^2}{A^a_b \partial
A^b_a}+\sum_nV_n{1\over N}\tr A^n \eeq where \m{V_n = N^{n-
1}\tilde V_n}. We are interested in determining the properties (
such as the ground state) of the system  in the limit as \m{N\to
\infty} keeping \m{V_n} and \m{\hbar} fixed.

We quote below the result obtained by Jevicki and Sakita
\cite{Collective2,Collective4,Collective5}, who changed variables
to the eigenvalue density $\rho(x)$ of \m{A} and showed that in the large
$N$ limit, the Hamiltonian may be written as
 \beqs
    {\tilde H \over N^2} &=& H = K_r + V_{coll} + \int dx ~V(x)
    ~\rho(x) \cr
    {\rm where~} V(x) &=& \sum_n V_n x^n.
 \eeqs
The collective potential and `radial' kinetic energy \footnote{We
use the term radial by analogy with atomic physics} are
 \beqs
    V_{coll} &=& \hbar^2{\pi^2 \over 6} \int ~ dx ~ \rho(x)^3
            = \hbar^2{\pi^2 \over 2} \int~ ((H\rho)(x))^2 ~ \rho(x)~ dx
        \cr
    K_r &=& \half \hbar^2\int~ dx~ \pi'(x)~ \rho(x)~ \pi'(x) \cr
    {\rm where~} \pi(x) &=& -{i \over N^2} {\delta \over \delta
    \rho(x)};~ [\pi(x),\rho(x')] = -{i \over N^2} \delta(x-x').
 \eeqs
We notice that their collective potential \m{V_{coll}} (the term that really
makes the Hamiltonian describing the large N limit different from
the finite N case) is just one fourth the free Fisher information
of one hermitian random variable.

This reformulation of the matrix model allows us to take the limit
\m{N\to\infty}  easily. Note that \m{N} only appears in the
commutation relations between \m{\rho} and \m{\pi}. Indeed, as
\m{N\to \infty} this commutator becomes small:\m{1\over N^2}
appears where \m{\hbar} appears in the Heisenberg relations. Thus
\m{N\to \infty} is a kind of classical limit. The limiting theory
has as configuration space the space of probability distributions
\m{\rho(x)} . The canonical conjugate of \m{\rho(x)} is \m{p(x)}:
\beq \{\rho(x),p(x')\} =  \delta(x-x') \eeq The hamiltonian \beq
H=\half \int~ dx~ \hbar^2p'^2(x)~ \rho(x)+V_{coll}(\rho)+\int
V(x)\rho(x)dx \eeq leads to the equations of motion \beqs {\pdr
p(x,t)\over \pdr t} &=& E-\half \hbar^2 p'^2(x)-{\pi^2\hbar^2\over
2}\rho(x)^2-V(x), \cr {\pdr \rho(x,t)\over \pdr t} &=& - \hbar^2
{\pdr\left[\rho(x)p'(x)\right]\over \pdr x} \eeqs where \m{E} is
the Lagrange multiplier imposing the constraint \m{\int
\rho(x)dx=1}. The spectrum (density of eigenvalues)
 of the matrix \m{A} in ground state of the matrix quantum mechanics is then
determined in this large N aproximation by the static solution
\beq
\rho(x)={1\over
\pi\hbar}\surd\left[2\left(E-V(x)\right)\right]\theta\left(E-V(x)\right).
\eeq
As noted above, the constant \m{E} is determined by the condition
\m{\int\rho(x)dx=1}.

Notice that the collective potential \m{V_{coll}}  plays a crucial
role in determining this answer; without that term there would
have been no such static (ground state) solution. The elimination
of the angular degrees of freedom induces a centrifugal potential
( collective potential \m{V_{coll}}) leading to a repulsion of
eigenvalues of \m{A}. The volume of the set of matrices of a given
spectrum vanishes when any pair of eigenvalues coincide. The
collective potential is the square of the derivative of the
logarithm of this volume, so is singular when a pair of
eigenvalues coincide. This repulsion of eigenvalues
counterbalances the tendency of the eigenvalues to accumulate at
the minimum of \m{V(x)}. It is this balance that leads to a
continuous
 density of eigenvalues.

 In the more conventional limit \m{\hbar\to 0} (keeping \m{N} fixed)  we
 would get a completely different answer: the ground state is concentrated at
 the minimum of \m{V(x)}.
\subsection{Several Degrees of Freedom}

Once we see the connection of the collective potential to the free Fisher
information,non-commutative probability theory immediately suggests the
generalization to multi-matrix models. (We will describe the answer first,
postponing its derivation to the next subsection.)The probability density \m{\rho(x)} no
longer makes sense with several random variables; instead we must think of the
moment tensors \m{G_I} as characterizing the non-commutative probability
distribution. This way of thinking leads to
 \beqs
    H &=& \half K_{ij} \pi^{ i I} G_{IJ}
        \pi^{j J  } + \ov{4} K_{ij} \Phi^{ij}(G) + V^I G_I \cr
    \pi^I &=& -{i \over N^2} \dd{}{G_I}; ~~ [\pi^I,G_J] =
        -{i \over N^2} \delta^{(J)}_{(I)}
    \label{formula-for-hamiltonian}
 \eeqs
where $\delta^{(J)}_{(I)} = 1$ if there is a cyclic permutation of
$I$ that equals $J$ and zero otherwise. We will now show by an
explicit change of variables that this is indeed the correct
hamiltonian of the large N limit. It is convenient to introduce
the effective potential

\beq
    V_{eff} = \ov{4} K_{ij} \Phi^{ij} + V^I G_I
\eeq

Thus, in the planar limit \m{N\to \infty} keeping \m{V^I, K_{ij}}
fixed we do get a classical mechanical system whose configuration
space is the space of non-commutative probability distributions
with Poisson brackets
 \beq
    \{G_I,\p^J \} = \delta^{(J)}_{(I)}
 \eeq
and the hamiltonian
\beqs
    H &=& \half K_{ij} p^{ i I} G_{IJ} p^{j J}
        \ + \ov{4} K_{ij} \Phi^{ij}(G) + V^I G_I
\eeqs
determine the equations of motion
 \beqs
    {dG_K \over dt} &=& \{ G_K,H \} = \{G_K,K_r\} \cr
        &=& \half K_{ij} \delta^{(J_1 j J_2)}_{(K)}
        \bigg[\pi^{I_1 i I_2} G_{I_2 I_1 J_2 J_1} +
        G_{I_2 I_1 J_2 J_1} \pi^{I_1 i I_2} \bigg] \cr
    {d\pi^K \over dt} &=& \{\pi^K,H \} \cr
        &=& \half K_{ij} \delta^{(K)}_{(I_2 I_1 J_2 J_1)}
        \pi^{I_1 i I_2} \pi^{J_1 j J_2} + \dd{V_{eff}(G)}{G_K}
 \eeqs
An interesting class of static solutions is determined by
\beqs
    \pi^{K} &=& 0 ~~ \forall~~ |K| \geq 1 \cr
    \dd{V_{eff}(G)}{G_K} &=& 0 ~~ \forall ~~ |K| \geq 1
 \eeqs

Thus, to find the ground state we must minimize $V_{eff}$. Note
that this variational principle involves no approximation aside
from the large $N$ limit. The equations for extremizing $V_{eff}$
are, $~L^k_K (\ov{4} K_{ij} \Phi^{ij} + V^I G_I ) = 0$. The chief
obstacle in solving these equations explicitly is the presence  of
the Fisher information term, which does not have any simple
expression in terms of the moments. Of the two forms of the
information matrix in eqn. (\ref{formula-for-information}), we
choose to use the one involving the auxiliary variables $S^I$'s in
carrying out the minimization of the effective potential. Let us
note here, that these variables $S^I(G)$ are to be thought of as
functions of the moments themselves, i.e. they are the solution to
the Schwinger-Dyson equations of some (static) matrix model, whose
large $N$ moments coincide with those of the dynamical matrix
model at a particular value of time. In terms of the auxiliary
variables $S^I(G)$ the equations to be solved are:
 \beq
    L^k_K V_{eff} = \ov{8} K_{ij} S^{jJ} \bigg[ 2 L^k_K \eta^i_J
            - S^{iI} L^k_K G_{IJ} \bigg] + V^{I_1 k I_2} G_{I_1 K
            I_2}= 0.
 \eeq

\subsection{A geometric derivation of the  collective Hamiltonian}

We will study the quantum mechanical system whose degrees of
freedom are described by a set of hermitian matrices \m{(A_i)^a_b}
and has hamiltonian \beq H  = -{1\over N^4}K_{ij}\half
\dd{^2}{(A_i)^a_b \pdr (A_j)^b_a} +
    {1\over N}\tr[ V^I A_I].
\eeq
The eigenvalue problem of \m{H} follows from the condition that the quadratic
form
\beq
    <\psi|H|\psi>=\int \left[
    {1\over 2 N^4}K_{ij}{\pdr\psi^*\over \pdr A_{ib}^a}{\pdr\psi\over \pdr A_{ja}^b}+{1\over
    N}\tr V^IA_I|\psi(A)|^2 \right]\prod_{i=1}^Md^{N^2}A_i
\eeq
be extremal, subject to the constraint on the norm \m{<\psi|\psi>=\int |\psi(A)|^2 dA=1}.

The kinetic energy term is just the laplacian with respect to a
  euclidean metric on the space \m{\cal A} of hermitean matrices:
 \beq
    \tilde g(d[A_i]^a_b,d[A_j]^c_d) = \tilde g^{~a~c}_{ibjd}
    = K_{ij} \delta^a_d \delta^c_b
 \eeq
This metric is invariant under the action of \m{U(N)} on the matrix variables:
 \m{A_i\to UA_iU^{-1}}. Moreover, the potential energy \m{\tr V^I A_I} is
 invariant under \m{U(N)}. We will regard this as  a `gauge transformation',
 and allow only for wavefunctions that are also invariant under \m{U(N)}:
 \beq
    \psi(UAU^\dag)=\psi(A).
 \eeq
 The problem is to determine the eigenvalues and eigenfunctions of the
 hamiltonian \m{H} over this class of invariant wave functions. Feynman diagram
 methods applied to this problem \cite{thooft} suggest that there is an
 important simplification in the  limit \m{N\to \infty} holding \m{V^I} and
 \m{K_{ij}} fixed: only Feynman diagrams of planar topology contribute. This in
 turn implies that the gauge invariant observables such as \m{G_I={1\over N}\tr
 A_I} have fluctutaions of the order \m{1\over N^2}: thus matrix models
 reformulated in terms of invariants tend to some classical theory in the large
 \m{N} limit. What are the configuration space, Poisson brackets and hamiltonian
 of this classical theory?

 A complete set of invariants for the matrices are given by \m{G_I={1\over N}\tr
 A_I}. For a fixed finite \m{N}, only a finite subset of the \m{G_I} can be independent
 variables. If we allow \m{N} to be arbitrary ( may be even infinite) the
 \m{G_I} are independent except for cyclic symmetry, the  reality conditions and inequalities.
 Thus the configuration space of matrix models is nothing but the space of
 non-commutative probability distributions that are tracial:
 \beq
 G_{\phi} =1,~~~ G_I^* = G_{\bar I}, ~~~ G_{IJ} {f^{\bar I}}^* f^J
 \geq 0 ~ {\rm for~any~polynomial~} f(X)
 \eeq
In essence our task is to change variables from cartesian co-ordinates
\m{A_{ib}^a} to `radial co-ordinates' \m{G_I}.

It is easier to change variables in the quadratic form \m{<\psi|H|\psi>}
 rather than the
hamiltonian operator \m{H} directly: we only have to deal with first order
rathar than second order derivatives. By the chain rule of differntiation,
\beq
    K_{ij}{\pdr\psi^*\over \pdr A_{ib}^a}{\pdr\psi\over \pdr A_{ja}^b}=
K_{ij}{\pdr G_I\over \pdr A_{ib}^a}{\pdr G_J\over \pdr A_{ja}^b}
{\pdr\psi^*\over \pdr G_I}{\pdr\psi\over \pdr G_J}.
\eeq
Now,
\beqs
    \dd{G_I}{[A_i]^a_b} &=&  \delta_I^{I_1 i I_2} [A_{I_2
        I_1}]^b_a \cr
    \Rightarrow \dd{G_I}{[A_i]^a_b} \dd{G_J}{[A_j]^b_a} &=& \delta_I^{I_1 i
        I_2} \delta_J^{J_1 j J_2} G_{I_2 I_1 J_2 J_1}
 \eeqs
 so that
 \beq
 K_{ij}{\pdr\psi^*\over \pdr A_{ib}^a}{\pdr\psi\over \pdr
 A_{ja}^b}=g_{I;J}{\pdr\psi^*\over \pdr G_I}{\pdr \psi\over \pdr G_J}
 \eeq
 where
 \beq
    g_{I;J} = K_{ij} \delta_I^{I_1 i I_2} \delta_J^{J_1 j J_2}
        G_{I_2 I_1 J_2 J_1}
 \eeq
 The geometrical meaning of this clear: \m{g_{I;J}} is the inverse of the
 metric tensor induced on the space of non-commutative probability
 distributions by the euclidean metric on the space of matrices.

 To complete the story we need to understand as well the Jacobian of the
 transformation \m{A_i\to G_I}. This problem was studied extensively in a
 previous paper, where we showed that this Jacobian is \m{e^{N^2\chi(G)}} where
 \m{\chi(G)} is the free entropy ( in the sense of Voiculescu) of the
 non-commutative probability distribution whose moments are \m{G_I}. Thus we get
 \beq
 <\psi|H|\psi>=\int\left[g_{I;J}{\pdr\psi^*\over N^2\pdr G_I}{\pdr \psi\over N^2\pdr
 G_J}+V^IG_I\psi^*\psi\right]e^{N^2\chi(G)}dG
 \eeq
 and, of course also,
 \beq
 <\psi|\psi>=\int \psi^*\psi e^{N^2\chi(G)}dG.
 \eeq
 To take the limit as \m{N\to \infty}, it is useful to introduce a new `radial
 wavefunction'
 \beq
    \Psi=e^{\half N^2\chi}\psi
 \eeq
 so that
 \beq
 <\psi|\psi>=\int \Psi^*\Psi dG,\quad  \int V^IG_I\psi^*\psi e^{N^2\chi(G)}dG=\int
 V^IG_I \Psi^*\Psi dG.
 \eeq

 In making this change of variables on the expectation value of kinetic energy,
 there will be some extra contributions from the derivatives of the
 entropy.These will add an extra term \m{V_{coll}} to the potential. It is our
 next task to determine this term. Using
 \beq
 {\pdr \psi\over N^2\pdr G_I}e^{\half N^2\chi}={\pdr \Psi\over N^2\pdr
 G_I}-\half {\pdr\chi\over \pdr G_I}\Psi
 \eeq
 we get
 \beqs
 g_{I;J}{\pdr\psi^*\over N^2\pdr G_I}{\pdr \psi\over N^2\pdr
 G_J}e^{N^2\chi(G)}&=&g_{I;J}{1\over N^2}
 {\pdr \Psi^*\over \pdr G_I}{1\over N^2}{\pdr \Psi\over \pdr G_J} +{1\over
 4}g_{I;J}{\pdr\chi\over \pdr G_I}{\pdr\chi\over \pdr G_J}|\Psi|^2\cr
 & & - {1\over 2} g_{I;J} {\pdr\chi\over \pdr G_I}{\pdr |\Psi|^2\over N^2\pdr G_J}
 \eeqs
 An integration by parts gives
 \beqs
 g_{I;J}{\pdr\psi^*\over N^2\pdr G_I}{\pdr \psi\over N^2\pdr
 G_J}e^{N^2\chi(G)}&=&g_{I;J}{1\over N^2}
 {\pdr \Psi^*\over \pdr G_I}{1\over N^2}{\pdr \Psi\over \pdr G_J} +{1\over
 4}g_{I;J}{\pdr\chi\over \pdr G_I}{\pdr\chi\over \pdr G_J}|\Psi|^2\cr
 & &  +{1\over 2} {1\over N^2}{\pdr \over \pdr G_J}\left[g_{I;J} {\pdr\chi\over \pdr
 G_I}\right] |\Psi|^2\cr
 & & +{\rm total\ divergence.}
 \eeqs
 The first term leads directly to the `radial' kinetic energy of
 eqn. (\ref{formula-for-hamiltonian}). The last two terms don't involve
 derivatives of \m{\Psi} and hence are
 contributions to the potential arising from the change of variables: they
 represent a sort of `centrifugal potential' ( or `collective potential' in the
 language of Jevicki-Sakita):
 \beq
 V_{coll}={1\over
 8}g_{I;J}{\pdr\chi\over \pdr G_I}{\pdr\chi\over \pdr G_J} +{1\over 4}
 {1\over N^2}{\pdr \over \pdr G_J}\left[g_{I;J} {\pdr\chi\over \pdr
 G_I}\right]
 \eeq
 In the large \m{N} limit the second term will be small compared to the
 first. In any case, we show that it vanishes. Using the formula for \m{g_{I;J}}
 and $\dd{\chi}{G_{Jj}} = G^{J;L} \eta^j_L$,

 \beqs
    \ov{4 N^2} \dd{}{G_J} (g_{I;J} ~\dd{\chi}{G_I})
        &=& \ov{4 N^2} K_{ij} \dd{}{G_I} \bigg( \delta_I^{I_1 i
        I_2} \delta_J^{J_1 j J_2} G_{I_2 I_1 J_2 J_1} \dd{\chi}{G_J}
                \bigg) \cr
     &=& \ov{4 N^2} K_{ij} \dd{}{G_{Ii}} \bigg(
        G_{IJ} \dd{\chi}{G_{Jj}} \bigg) = \ov{4 N^2} K_{ij}
        \dd{\eta^j_I}{G_{Ii}} \cr
    &=& \ov{4 N^2} K_{ij} \dd{}{G_{Ii}}
        \bigg(\delta_I^{I_1 j I_2} G_{I_1} G_{I_2} \bigg) = 0.
 \eeqs
since we are always differentiating with respect to a higher
moment. Thus only the first term contributes to $V_{coll}$
 \beqs
 V_{coll} &=& {1\over 8} K_{ij}G_{I_2 I_1 J_2 J_1}
 {\pdr\chi\over \pdr G_{I_1iI_2}}{\pdr\chi\over \pdr G_{J_1jJ_2}}
     \cr
    &=& \ov{8} K_{ij} G_{I_2 I_1 J_2 J_1} G^{I_2 I_1;L} \eta^i_L
    G^{J_2 J_1; M} \eta_M^j \cr
    &=& \ov{8} K_{ij} G^{L;M} \eta^i_L \eta^j_M
 \eeqs
 From our earlier discussion of the Fisher information we see that
 \beq
 V_{coll}={1 \over 4}K_{ij}\Phi^{ij}.
 \eeq

 Now we can collect the different terms together to get the answer for the
 hamiltonian stated in eqn. (\ref{formula-for-hamiltonian}).

\section{Approximate Large N Ground State and the Variational Principle}

Our aim is to understand the ground state of the Hamiltonian
 \beq
    H = K_r + V_{eff}
 \eeq
where $K_r = \half K_{ij} \pi^{I_1 i I_2} G_{I_2 I_1 J_2 J_1}
\pi^{J_1 j J_2}$ and $V_{eff} = \ov{4} K_{ij} \Phi^{ij} + V^I
G_I$. The large $N$ ground state is the probability distribution
minimizing the effective potential $V_{eff} = V_{coll} + V^I G_I$.
In essence, this is the problem of minimizing free Fisher
information holding the moments in the potential fixed via the
Lagrange multipliers $V^I$. So far we have not made any
approximations aside from the large $N$ limit.

{\bf{Mean Field Theory:}} The equation for minimization of the
effective potential also defines for us a variational principle.
If we minimize $V_{eff}$ over a sub-manifold of the configuration
space which corresponds to the choice of a few parameter family of
actions for the static matrix model, we will get an upper bound
for the ground state energy. But as we well know, explicit
solutions of the Dyson-Schwinger equations for arbitrary
multi-matrix models are not always possible. However we may chose
as the action $S$ some solvable model, which can then act as a
variational ansatz. The simplest choice for the action is the
gaussian, which leads to the wignerian sub-manifold of moments.
This leads to an approximation method similar to mean field theory
(MFT). As a first check on our methods, we  present below
mean-field calculations for some dynamical matrix models.

Since the gaussian action for static matrix models leads to Wigner
moments,
 \beq
    \Phi^{ij}(G) = \half S^{iI}(G) S^{jJ}(G) G_{IJ} = \half (G^{-1})^{ij}
 \eeq
on the wignerian sub-manifold. Here $(G^{-1})^{ij} G_{jk} =
\delta^i_k$. Let us specialize to matrix models with a quartic
potential $V^{ij} G_{ij} + V^{ijkl} G_{ijkl}$. Then the effective
potential to be minimized is

\beq
    V_{eff}(G_{ij}) = \ov{8}K_{ij} (G^{-1})^{ij} + V^{ij} G_{ij}
        + V^{ijkl} (G_{ij} G_{kl} + G_{il} G_{jk}).
 \eeq
The minimum of the effective potential is given by,
 \beqs
    \dd{V_{eff}}{G_{pq}} &=& -\ov{8} K_{ij} (G^{-1})^{ip}
        (G^{-1})^{jq} + V^{pq} \cr && + V^{pqkl} G_{kl}
        + V^{ijpq} G_{ij} + V^{pjkq} G_{jk} + V^{ipql} G_{il} = 0.
 \eeqs
Notice that this is a cubically non-linear equation for the
positive symmetric variational matrix $G_{ij}$. $G_{ij}$ plays the
role of a mean field and this is its self consistent equation.

\subsection{Mean Field Theory for Dynamical $1$ Matrix Model}

To calibrate our approximation method, let us consider the dynamical
$1$ matrix model
 \beq
    H = -\half \dd{^2}{A^a_b \pdr A^b_a} + \tr [\half A^2 + g A^4]
 \eeq
whose exact large $N$ ground state energy is known from the work
of Brezin et.al. \cite{brezin-et-al}.
 On the sub-manifold of Wigner semicircular distributions,
 \beq
    V_{eff}^{sc}(G_2) = \ov{8 G_2} + \half G_2 + 2g G_2^2
 \eeq
whose minimum occurs at
 \beqs
    G_2^{sc}(g) &=& \ov{24g}[-1 + \mu^{-1/3} +
        \mu^{1/3}], \cr
    {\rm where~~} \mu &=& -1 + 216 g^2
        + 12g \sqrt{-3 + 324 g^2}
 \eeqs
The semicircular estimate is a variational upper bound $E_{gs}
\leq E_{gs}^{sc}$.

\subsection{Cram$\acute{\rm e}$r-Rao Inequality}

Somewhat surprisingly, we
can also get a lower bound for
$E_{gs}$. For this we make use of the (non-commutative)
Cram$\acute{\rm e}$r-Rao (CR) inequality
\cite{Rao, Pit, Voic4} along with the moment
inequalities, which are conditions for positivity of the Hankel
matrix. The CR inequality states that $\Phi^{ij} - (G^{-1})^{ij}$
is a positive matrix. It is the statement that among all
distributions with fixed 2nd moment, the Wigner distribution
minimizes information. For one variable, it is the condition
$\Phi~ G_2 \geq 1$ which is understood as a sort of uncertainty
principle in statistics. The moment inequalities (MI) for a single
variable are
 \beq
    G_2 \geq 0, ~~~ G_4 \geq G_2^2,~~~
    G_2 G_4 G_6 - G_4^3 + G_2^3 G_6 - G_2^2 G_4^2 \geq 0,~~~ \cdots
 \eeq
We can saturate the MI by choosing $G_2 \geq 0;~~ G_4 = G_2^2;~~
G_6 = G_2^3 \cdots$. The MI are saturated by $\rho(x) =
\delta(x-x_0)$, $G_{2n} = x_0^{2n}$, {\it not} the Wigner
moments.

The CR-MI estimate is a lower bound $E_{gs}^{CR-MI} \leq E_{gs}$.
We see from the following comparison with Brezin et.al.'s exact
($N \to \infty$) result, that in addition to providing bounds on
$E_{gs}$, mean field theory is a reasonably good approximation.

\begin{tabular}{|c|c|c|c|} \hline
  g & $E_{gs}^{CR-MI}$ & $E_{gs}^{exact}$ & $E_{gs}^{sc}$ \\ \hline \hline
  .01 & .503 & .505 & .505 \\ \hline
  .1 & .523 & .542 & .543 \\ \hline
  .5 & .591 & .651 & .653 \\ \hline
  1 & .653 & .740 & .743 \\ \hline
  50 & 1.794 & 2.217 & 2.236 \\ \hline
  1000 & 4.745 & 5.915 & 5.969 \\ \hline
  $g \to \infty$ & .472 $g^{1/3}$ & .590 $g^{1/3}$ & .595 $g^{1/3}$ \\
\hline
\end{tabular}

\subsection{Semi-Circular Ansatz for a Dynamical $2$ Matrix Model}

We consider the two matrix model, that is of considerable interest in
M-theory \cite{BFSS}(for simplicity we look at the non-supersymmetric
counterpart of the `BFSS model), defined by,
 \beqs
    H &=& -\half \bigg(\dd{^2}{A^b_a \pdr A^a_b} + \dd{^2}{B^b_a \pdr
            B^a_b} \bigg) + V(A,B) \cr
    {\rm where ~~} V(A,B) &=& \tr \bigg[ {m^2 \over 2}(A^2 + B^2)
        + {c \over 2} (AB + BA) - {g \over 4}[A,B]^2 \bigg].
 \eeqs
We read off $V^{ij} = \pmatrix{ \half m^2 & \half c \cr \half c &
\half m^2}$, $V^{ABAB}=-V^{ABBA} = -\half g$ and consider the
region $g \geq 0,~|c| \leq m^2$ where $V^{ij}$ is positive. The
effective potential to be minimized is
 \beq
    V_{eff} = \ov{8} \tr [G^{-1}] + \half m^2 (G_{AA}+G_{BB})
            + c G_{AB} - \half g (G_{ABAB} - G_{ABBA})
 \eeq
In terms of the positive $G_{ij} = \pmatrix{\alpha & \beta \cr
\beta & \alpha}$ ($|\beta| \leq \alpha$) parameterizing the
Wignerian sub-manifold of configuration space,
 \beq
    V_{eff}(\alpha,\beta) = {\alpha \over 4(\alpha^2 - \beta^2)}
        + m^2 \alpha + c \beta + \half g (\alpha^2 - \beta^2).
 \eeq
For the above range of parameters, $V_{eff}$ is bounded below.
Since the information and potential scale oppositely, it has a
minimum at a non-trivial $\alpha > 0$. For $c = 0$, we get $\beta
= 0$ and a cubic for $\alpha: 4g\alpha^3 + 4 m^2 \alpha^2 -1=0$.
For $c>0$ the actual solution of the algebraic equations can be
obtained numerically.

\subsection{MFT for Relativistic $\la \phi^4$ Matrix Field Theory}

Next we
consider a relativistic hermitian scalar matrix field
$\phi^a_b(x)$ in $d+1$ dimensions. The Hamiltonian in the
Schrodinger representation is
 \beq
    H = \int d^d x \bigg[ -\half {\delta^2 \over \delta \phi^a_b(x)
         \delta \phi^b_a(x)} + \tr \bigg(\half |\grad \phi(x)|^2
         + {m_0^2 \over 2} \phi(x)^2 + \la_0 \phi(x)^4 \bigg) \bigg].
 \eeq
Let $G_{xy} = \Ntr \phi(x) \phi(y)$, $\int d^dy G^{xy} G_{yz} =
\delta^d(x-z)$. Within the mean field approximation, the effective
potential is
 \beq
    V_{eff} = \ov{8} \int dx G^{xx} + \int dx dy \half
    \delta(x-y) (-\dd{^2}{x^2} + m_0^2) G_{xy} + 2\la_0 \int dx (G_{xx})^2.
 \eeq
Minimizing it leads to an integral equation for the mean field
$G_{xy}$
\beq
    -\ov{8} \delta(x-y) + \half \int dz G_{zx} (-\dd{^2}{z^2} +
    m_0^2)G_{zy} + 4\la_0 \int dz G_{zz} G_{zx} G_{zy} =0
 \eeq
Assuming a translation invariant ground state
 \beq
    G_{xy} = G(x-y) = \int {d^dp \over (2\pi)^d} e^{ip(x-y)}
        \tilde G(p),
 \eeq
this equation can be solved:
 \beq
    \tilde G(p) = \ov{2\sqrt{p^2 + m^2}}
 \eeq
where $m$ is determined self consistently by
 \beq
    m^2 = m_0^2 + 4\la_0 \int^\La {d^d p \over (2\pi)^d} \ov{\sqrt{p^2 +
m^2}}
 \eeq
We recognize this as the mass renormalization; log, linearly and
quadratically divergent in 1, 2 and 3 spatial dimensions. We need
to go beyond the mean field ansatz to see coupling constant
renormalization. We will address these issues in a separate paper.

Acknowledgement: \DOE

% \bibliography{abhishekbib}

\end{document}